\documentclass[12pt,a4pape,showkeys]{revtex4}
\usepackage{graphicx}
\usepackage{epsfig}
\usepackage{bm}
\usepackage{amsmath}
\usepackage{amssymb}
\usepackage{graphics}
\usepackage{epstopdf}
\usepackage[linkcolor=red,citecolor=green,urlcolor=blue]{hyperref}

\begin{document}

\title{ \textbf{On the QCD Evolution of Transverse Momentum Dependent Distributions}}
\author{Federico Alberto Ceccopieri}
\email{federico.alberto.ceccopieri@cern.ch}
\affiliation{IFPA, Universit\'e de Li\`ege,  All\'ee du 6 ao\^ut, B\^at B5a, 4000
Li\`ege, Belgium}
\author{Luca Trentadue}
\email{luca.trentadue@cern.ch}
\affiliation{Dipartimento di Fisica e Scienze della Terra ``Macedonio Melloni'', Universit\'a di Parma and INFN,
Sezione di Milano Bicocca, Milano, Italy.}
\begin{abstract}
\noindent
We reconsider the evolution equations for transverse momentum dependent distributions 
recently proposed by us and recast them in a form which allows the comparison with results recently appeared in the literature.  We show under which conditions the obtained results might be consistent with each other.
\end{abstract}

\maketitle

\section{Introduction}
\noindent
Transverse momentum dependent (TMD) distributions are currently object of intense research activity due to 
their wide range of applicability in many areas of QCD, ranging, for example, 
from small-$x$ physics to spin physics. The use of TMD distributions is indeed phenomenologically 
appealing since, as it is known from a long time, observables constructed upon them show a reasonable 
agreement with data already to lowest order in the perturbative expansion, 
which is not the case for predictions based on collinear factorization at the same accuracy. 
The concept itself of TMD distribution is closely connected to the description 
of QCD hard processes at small transverse momentum, being 
the $p_t$-spectrum of gauge bosons produced in hadronic collisions a well known example.
At low $p_t$ the latter manifests many perturbative and non-perturbative features of the underlying theory.
In particular techniques for the resummation of the perturbative series in the multiple soft gluon emission limit 
were first developed for this prototype observable~\cite{DDT,PP,CSS} and 
for the closely connected case of dihadron production at small relative $p_t$ in 
$e^+ e^-$ annihilation~\cite{BCM,KT1,KT2,CS}. 
More recently these techniques have been also extended 
the relevant case of Higgs production in hadronic collisions~\cite{Higgs0,Higgs} and 
to processes containing coloured final states, which indeed add non trivial issues to the resummation formalism. We mention, as representative examples, the production of heavy quarks~\cite{SGR_HQ},
of prompt photon~\cite{prompt_photon}, of dijet~\cite{SGR_dijet}, of top-pair~\cite{top-pair}, of single and double~\cite{SGR_1_particle,dihadron} hadrons in hadronic collisions as well as single inclusive hadron production in Semi-Inclusive Deep Inelastic Scattering~\cite{SGR_SIDIS}.
Resummation techniques have also been applied to the hadronic transverse energy flow associated 
with weak and Higgs bosons production in hadronic collisions~\cite{TEF_2010,TEF_2014} which might represent a valuable observable for the characterisation of the underlying event. 
A very successful theoretical and phenomenological program 
has been developed for all these processes and the state-of-the-art analyses in this field will likely
improve the accuracy of resummed perturbative calculations, see for example the recent results presented
in Refs.~\cite{CCDFG1,CCDFG2}. 

On the other hand, individual TMD distributions, which have been historically the starting 
point of the resummation program discussed above, have not progressed at same rate.
This is mainly due to non trivial obstacles in the proper QCD definition and evolution of these distributions
and the related and important issue of TMD factorisation of the relevant cross sections.   
In particular, the precise knowledge of QCD evolution 
of TMD distributions would allow a combined description of data coming from 
experiments at different energies. This in turn would allow to test TMD factorisation quantitatively
and therefore to constrain the non-perturbative part of TMD distributions~\cite{BLNY,Melis,SY}.

In the recent past a number of theoretical analyses have appeared in the literature
focusing both on TMD factorisation and evolution~\cite{JMY,Stefanis,Hautmann}. 
Quite recently a formal definition of TMD distributions has been given in Ref.~\cite{JC}.
The QCD evolution for TMDs based on such definition have been presented in Ref.~\cite{AR}\,.
Nearly on the same time these issues have been addressed by another group independently, see for example Refs.~\cite{EIS1,EIS2}. The compatibility of these last two approaches has been discussed in Ref.~\cite{CR}
and it was further established in Ref.~\cite{EIS3}.

The aim of the present note is to compare our formalism, originally proposed in Refs.~\cite{BCM,KT1,CT}, with the 
ones mentioned above. In order to compare the predicted structure of QCD evolution for 
TMD distributions, we will focus on the pure perturbative contributions as predicted by the various formalisms. 
At the same time, since our interest is in the identification of the eventually common structures, 
the comparison will be carried out at leading (and partially at the subleading) logarithmic accuracy, 
althought this approximation is rather out-of-date given the accuracy reached in state-of-the-art analyses. 

The paper is organised as follows. In Sec.~\ref{my} we briefly review the evolutions equations
for TMD distributions obtained by us and recast them in a form which
will facilitate further comparisons. In Sec.~\ref{secAR} and Sec.~\ref{secEIS} we report and elaborate
the relevant formulas from the other two approaches we want to compare to and discuss 
the level of agreement between these results focusing on the Drell-Yan as reference process.
We collect our conclusions in Sec.~\ref{Conc}.   

\section{Unpolarised evolution}
\label{my}
\noindent
The evolution equations for transverse momentum distributions were originally derived in 
Ref.~\cite{BCM} in timelike kinematics, \textsl{i.e.} for unpolarised TMD
fragmentation functions. Later on the same evolution equations were used to perform 
the resummation of leading and subleading logarithmic corrections to nearly back-to-back hadrons 
produced in $e^+ e^-$ annihilation for the so called energy-energy correlation observable~\cite{KT1,KT2}.
More recently they have been generalised to spacelike kinematics, \textsl{i.e.} for 
unpolarised TMD parton distribution functions and fracture functions~\cite{CT}. 
Such an extension essentially requires only 
minor changes in parton kinematics. In this case the evolution equations read:
\begin{multline}
\mu^2 \frac{\partial}{\partial \mu^2} f_{i/P}(x,\bm{k_{\perp}},\mu^2)=\frac{\alpha_s(\mu^2)}{2\pi}
\int_x^1 \frac{du}{u^3} P_{ji}(u) 
\int \frac{d^2 \bm{l_{\perp}} }{\pi} \cdot \\ \cdot
\delta[l_{\perp}^2-(1-u)\mu^2] f_{j/P}\Big(\frac{x}{u},\frac{\bm{k_{\perp}}-\bm{l_{\perp}}}{u},\mu^2\Big)\,.
\label{eq::1}
\end{multline}
TMD parton distribution functions $f_{i/P}(x,\bm{k_{\perp}},\mu^2)$ in eq.~(\ref{eq::1}) give
the probability to find, at a given scale $\mu^2$, a parton  $i$ 
with fractional momentum $x$ and transverse momentum $\bm{k}_{\perp}$ 
relative to the parent hadron. $P_{ji}(u)$ are the spacelike splitting functions~\cite{DGLAP}.
In order to better clarify the physical content of the equations let us consider 
a spacelike parton cascade. At each branching, as a consequence of parton radiation, the active parton increases its virtuality and acquires a small transverse momentum with respect to the parent.
These iterated emissions  generate therefore an appreciable transverse momentum, 
up to the order of the hard scale in the process, which adds to the non-perturbative one due to  
Fermi motion of the parton in the parent hadron.
Collinear emissions contribute large logarithmic corrections 
when the transverse momenta are ordered along the ladder and can be resummed to all orders
by using DGLAP evolution equations~\cite{DGLAP}. 
In the unintegrated case, the integration on relative transverse momenta, $\bm{l_{\perp}}$, generated between the two daughter partons at each branching, is left undone. In particular, we may consider one of these 
branchings, namely $p_i(\tilde{k}) \rightarrow p_j(k)+p_k(l)$, where we have indicated four-momenta in parenthesis.
With this notation the following mass-invariant constraint at the branching vertex can be derivered:
\begin{equation}
\label{mass}
l^2_{\perp}=-(1-u)k^2+u(1-u)\widetilde{k}^2-u l^2\,,
\end{equation}
where $u$ is the splitting variable. 
If one assumes that the virtualities increase along the ladder, $k^2\gg\widetilde{k}^2$, 
and on-shell partons are emitted,  $l^2=0$, the last two terms in eq.~(\ref{mass})
can be disregarded. In these limits, setting $-k^2=\mu^2$, 
the argument of the $\delta$-function in eq.~(\ref{eq::1}) is obtained.  
The transverse arguments of $f_j$ on r.h.s. of eq.~(\ref{eq::1}) are derivered 
by taking into account the Lorentz boost of transverse momenta from the emitting parton 
$\widetilde{k}$ reference frame to the interacting parton $k$ one~\cite{GSW}. 
In particular, the transverse momentum $\bm{\widetilde{k}}_{\perp}$ of the parton which undergoes the splitting
can be expressed as follows 
\begin{equation}
\label{boost}
\bm{\widetilde{k}}_{\perp}=(\bm{k}_{\perp}-\bm{l}_{\perp})/u\,.
\end{equation}
The latter is in fact the major change with respect to evolution equations for timelike kinematics. 
We assume that, upon integration over $\bm{k_{\perp}}$, the TMD parton distribution functions $f_{i/P}(x,\bm{k_{\perp}},\mu^2)$ reduce to their collinear conterparts
\begin{equation}
\int d^2 \bm{k_{\perp}}  f_{i/P}(x,\bm{k_{\perp}},\mu^2)= f_{i/P}(x,\mu^2)\,.
\end{equation} 
It can be shown that this normalisation condition is fulfilled by the evolution equations 
in eq.~(\ref{eq::1}), since, upon integration over $\bm{k_{\perp}}$, they reduce to collinear
DGLAP evolution equations~\cite{DGLAP}. Such normalisation condition may serve as a powerful check of the evolution. These equations can be solved numerically directly in transverse momentum space once suitable 
initial conditions for $f_{i/P}(x,\bm{k_{\perp}},Q_0^2)$ are provided, being $Q_0^2$ the starting scale 
for the evolution. 
They have been used to compute the $p_t$-spectrum charged hadron production in Semi-Inclusive 
Deep Inelastic Scattering in Ref.~\cite{CT2} and the $p_t$-spectrum of weak boson
produced in hadronic collisions in Ref.~\cite{Zpt_Ceccopieri}. On more formal grounds they have been used 
recently to investigate a number of sum rules for TMD distributions in Ref.~\cite{RT}.

The proposed evolution equations however do not reproduce the structure 
of leading (double) logarithms which show up in every fixed order calculation in perturbation
theory at small transverse momemtum. In order to correctly reproduce such terms, the argument of the running coupling is taken to be as the relative transverse momentum $l_\perp^2=(1-u)\mu^2$ at each parton branching~\cite{DDT,PP}
so that we define the modified splitting function 
\begin{equation}
\overline{P}_{qq}(u,\mu^2)=\bigg[ C_F \frac{1+u^2}{1-u} \frac{\alpha_s((1-u)\mu^2)}{2\pi} \bigg]_+\,.
\label{eq::2b}
\end{equation}
Upon expansion of the strong coupling around $u\simeq 0$, it is then easy to show that this replacement
amount to the resummation of a whole tower of large logarithms:
\begin{equation}
\overline{P}_{qq}(u,\mu^2)=C_F \frac{\alpha_s(\mu^2)}{2\pi}\bigg[ \frac{1+u^2}{1-u} \bigg]_+ 
- \beta_0 C_F \frac{\alpha^2_s(\mu^2)}{2\pi}\bigg[ \frac{1+u^2}{1-u} \, \ln(1-u) \bigg]_+ +\cdots
\label{eq::2c}
\end{equation}
In this expansion the first term on the right hand side corresponds to the usual Altarelli-Parisi
kernel whereas the second one corresponds to the first logarithmic enhanced term due to 
incomplete cancellation of contributions from real-virtual parton radiation. 
Since dominant corrections do appear in the quark-to-quark channel it is sufficient 
to consider the non-singlet component of the evolution equations. Analogous results can be derivered in the gluon  sector~\cite{CDT}, relevant, among others, for the description of the Higgs boson $p_t$-spectrum.
With these replacements the evolution equations in eq.~(\ref{eq::1}) read
\begin{equation}
\mu^2 \frac{\partial}{\partial \mu^2} f_{ns}(x,\bm{k_{\perp}},\mu^2)=
\int_x^1 \frac{du}{u^3} \overline{P}_{qq}(u,\mu^2) 
\int \frac{d^2 \bm{l_{\perp}} }{\pi}
\delta[l_{\perp}^2-(1-u)\mu^2]f_{ns}\Big(\frac{x}{u},\frac{\bm{k_{\perp}}-\bm{l_{\perp}}}{u},\mu^2\Big)\,.
\label{eq::2}
\end{equation}
In what follows we essentially repeat the calculation of Refs.~\cite{KT1,KT2} but for spacelike
evolution equations. The final result will be unchanged with respect to the one presented in Refs.~\cite{KT1,KT2}, since in the particular limit we will consider, the resummation of logarithmic enhanced terms is not affected by kinematics. 
We then introduce the two-dimensional Fourier-trasform of transverse momentum distributions defined by
\begin{equation}
\mathcal{F}_{ns}(x,\bm{b_{\perp}},\mu^2)=
\int d^2 \bm{k_{\perp}} e^{-i \bm{b_{\perp}} \cdot \bm{k_{\perp}}}f_{ns}(x,\bm{k_{\perp}},\mu^2)\,,
\label{eq::3}
\end{equation}
where the transverse vector $\bm{b_{\perp}}$ is the Fourier-conjugated of $\bm{k_{\perp}}$.
Applying this transformation to the r.h.s. of eq.~(\ref{eq::2}) we get
\begin{equation}
\int d^2 \bm{k_{\perp}} e^{-i \bm{b_{\perp}} \cdot \bm{k_{\perp}}} f_{ns}\Big(\frac{x}{u},\frac{\bm{k_{\perp}}-\bm{l_{\perp}}}{u},Q\mu^2\Big)=u^2 e^{-i \bm{b_{\perp}} \cdot \bm{l_{\perp}}}
\mathcal{F}_{ns}\Big(\frac{x}{u},u\bm{b_{\perp}},\mu^2\Big)\,,
\label{eq::4}
\end{equation}
so that the transformed equation in $b$-space reads
\begin{multline}
\mu^2 \frac{\partial}{\partial \mu^2} \mathcal{F}_{ns}(x,\bm{b_{\perp}},\mu^2)=
\int_x^1 \frac{du}{u} \overline{P}_{qq}(u,\mu^2)
\int \frac{d^2 \bm{l_{\perp}} }{\pi} \cdot \\ \cdot
\delta[l_{\perp}^2-(1-u)\mu^2] e^{-i \bm{b_{\perp}} \cdot \bm{l_{\perp}}}
\mathcal{F}_{ns}\Big(\frac{x}{u},u\bm{b_{\perp}},\mu^2\Big)\,.
\label{eq::5}
\end{multline}
Assuming that $\mathcal{F}_{ns}$ does not depend upon the azimuthal angle, by rotational invariance, the angular part of the two-dimensional integral in $\bm{l_{\perp}}$ can be expressed 
in terms of the Bessel function of the first kind, $J_0$, defined by
\begin{equation}
J_0(z)=\frac{1}{\pi}\int_0^\pi d\theta \, e^{i z \cos\theta}\,, 
\label{eq::5b}
\end{equation}
where in our case $z=|\bm{b_{\perp}}||\bm{l_{\perp}}| \equiv bl$ and $\theta$ is the relative angle
between transverse vectors $\bm{b_{\perp}}$ and $\bm{l_{\perp}}$ in transverse plane.
With these changes the equation becomes:
\begin{equation}
\mu^2 \frac{\partial}{\partial \mu^2} \mathcal{F}_{ns}(x,b,\mu^2)=
\int_x^1 \frac{du}{u} \overline{P}_{qq}(u,\mu^2) \int dl^2 \delta[l^2-(1-u)\mu^2] J_0(bl)
\mathcal{F}_{ns}\Big(\frac{x}{u},ub,\mu^2\Big)\,.
\label{eq::6}
\end{equation}
The last integral can be easily evaluated with the help of the $\delta$-function. 
As already stated, leading logarithmic corrections arise in the so called soft limit,
\textsl{i.e.} when emitted partons (in the present case, gluons) have vanishing energy. 
We set therefore $u=1$ in all slowly varying terms appearing in the evolution equation.
In such a limit, the equation is easily solved and a general solution is given by
\begin{equation}
\mathcal{F}_{ns}(x,b,Q^2)= \mathcal{F}_{ns}(x,b,Q_0^2) \exp[T(Q_0^2,Q^2,b)]\,,
\label{eq::8}
\end{equation}
where $Q_0^2$ and $Q^2$ are respectively the initial and final scale of the evolution and 
both must provided in the perturbative regime. 
The exponent of the quark form factor, $T(Q_0^2,Q^2,b)$, is defined by
\begin{equation}
T(Q_0^2,Q^2,b)=
\int_{Q_0^2}^{Q^2}\frac{d\mu^2}{\mu^2} \int_x^1 du \bigg[ \frac{\alpha_s((1-u)\mu^2)}{2\pi} \widehat{P}_{qq}(u)\bigg]_+ J_0(b\sqrt{(1-u)\mu^2})\,,
\label{eq::9}
\end{equation}
where the unregularised splitting function $\widehat{P}_{qq}$ is given by
\begin{equation}
\widehat{P}_{qq}(u)=C_F \frac{1+u^2}{1-u}=C_F \bigg(\frac{2}{1-u}-(1+u) \bigg)\,,
\label{eq::9b}
\end{equation}
and the last equality is displayed for later convenience. 
Applying the definition of the $+$-distribution we get 
\begin{multline}
T(Q_0^2,Q^2,b)=
\int_{Q_0^2}^{Q^2} \frac{d\mu^2}{\mu^2} \Bigg[ \int_x^1 du \frac{\alpha_s((1-u)\mu^2)}{2\pi} \widehat{P}_{qq}(u) J_0(b\sqrt{(1-u)\mu^2}) + \\ - 
\int_0^1 du \frac{\alpha_s((1-u)\mu^2)}{2\pi} \widehat{P}_{qq}(u) \Bigg]\,, 
\label{eq::10}
\end{multline}
where the first and the second term represent respectively the real and virtual emission terms and they are separately 
divergent in the $u\rightarrow1$ limit. It is then useful to rearrange such integral as
\begin{multline}
T(Q_0^2,Q^2)=
\int_{Q_0^2}^{Q^2} \frac{d\mu^2}{\mu^2} \Bigg[ \int_x^1 du \frac{\alpha_s((1-u)\mu^2)}{2\pi} \widehat{P}_{qq}(u) [J_0(b\sqrt{(1-u)\mu^2})-1] + \\ - \int_0^x du \frac{\alpha_s((1-u)\mu^2)}{2\pi} \widehat{P}_{qq}(u) \Bigg]\,. 
\label{eq::11}
\end{multline}
In this form the cancellation of real-virtual divergences in the soft limit is manifest since 
for $u \rightarrow 1$ one has $J_0(0)=1$. We also note that in such limit 
the strong coupling constant is evaluated in the infrared. Therefore a regularisation 
procedure must be eventually provided, a fact which is common to all resummation procedure. 
Changing integration variable from $u$ to $q^2=(1-u)\mu^2$ (note that $q^2$ is not related to $Q^2$ 
in eq.~(\ref{eq::8})), we get
\begin{multline}
T(Q_0^2,Q^2,b)=
-\frac{C_F}{\pi}\int_{Q_0^2}^{Q^2} \frac{d\mu^2}{\mu^2} \Bigg[ \int_0^{c\mu^2} \frac{dq^2}{q^2} \alpha_s(q^2) \bigg(1-\frac{q^2}{\mu^2}+\frac{q^4}{2\mu^4}\bigg) [1-J_0(bq)] + \\+ \int_{c\mu^2}^{\mu^2} \frac{dq^2}{q^2} \alpha_s(q^2)\bigg(1-\frac{q^2}{\mu^2}+\frac{q^4}{2\mu^4}\bigg) \Bigg]\,,
\label{eq::12}
\end{multline}
with $c=1-x$. It is now useful to write
\begin{equation}
1-J_0(bq)=\theta(bq-1)+R(bq)\,,
\label{eq::13}
\end{equation}
where $R(bq)$ is a finite reminder function. Eq.~(\ref{eq::12}) then becomes
\begin{eqnarray}
T(Q_0^2,Q^2,b)&=&
-\frac{C_F}{\pi}\int_{Q_0^2}^{Q^2} \frac{d\mu^2}{\mu^2} \int_0^{c\mu^2} \frac{dq^2}{q^2} \alpha_s(q^2) \bigg[1-\frac{q^2}{\mu^2}+\frac{q^4}{2\mu^4}\bigg] \theta(bq-1) + \nonumber\\
&&-\frac{C_F}{\pi}\int_{Q_0^2}^{Q^2} \frac{d\mu^2}{\mu^2} +\int_{c\mu^2}^{\mu^2} \frac{dq^2}{q^2} \alpha_s(q^2)\bigg[1-\frac{q^2}{\mu^2}+\frac{q^4}{2\mu^4}\bigg] + \nonumber\\
&&-\frac{C_F}{\pi}\int_{Q_0^2}^{Q^2} \frac{d\mu^2}{\mu^2} \int_0^{c\mu^2} \frac{dq^2}{q^2} \alpha_s(q^2) \bigg[1-\frac{q^2}{\mu^2}+\frac{q^4}{2\mu^4}\bigg] R(bq)\,.
\label{eq::14}
\end{eqnarray}
The first two terms, taking into account the constraint of the $\theta$-function, can be added together
to give  
\begin{equation}
T_{A}(Q_0^2,Q^2,b)=
-\frac{C_F}{\pi}\int_{Q_0^2}^{Q^2} \frac{d\mu^2}{\mu^2} \int_{1/b^2}^{\mu^2} \frac{dq^2}{q^2} \alpha_s(q^2) \bigg[1-\frac{q^2}{\mu^2}+\frac{q^4}{2\mu^4}\bigg]\,. 
\label{eq::15}
\end{equation}
At this point is useful to consider the hierarchy of scales of the problem. 
In particular we suppose to hold the following inequality $Q_0^2<b_0^2/b^2<Q^2$, \textsl{i.e.} 
the transverse momentum $b_0^2/b^2$ (with $b_0$ an arbitrary constant) is always greater that the infrared cut-off $Q_0^2$ and smaller that $Q^2$: the case $b_0^2/b^2 \gg Q^2$ corresponds infact to the emission of hard 
gluons and can be treated in fixed order perturbation theory. 
However if $b_0^2/b^2$ is much larger than $Q_0^2$ (but still less than $Q^2$), large logarithms of the type $\ln(Q_0^2 b^2/b_0^2)$ will appear in the final result. In order to avoid such terms 
and optimise the perturbative expansion,  we may therefore 
set $Q_0^2=b_0^2/b^2$. In this case changing the order of integration in eq.~(\ref{eq::15}) and integrating over $\mu^2$ we get 
\begin{equation}
T_{A}(b_0^2/b^2,Q^2)=
-\frac{C_F}{\pi} \int_{b_0^2/b^2}^{Q^2} \frac{dq^2}{q^2} \alpha_s(q^2) \bigg[ \ln\frac{Q^2}{q^2}
-1+\frac{q^2}{Q^2}+\frac{1}{4}-\frac{q^4}{4Q^4} \bigg]\,.
\label{eq::16}
\end{equation}
Disregarding term which give upon integration power suppressed terms we obtain
\begin{equation}
T_{A}(b_0^2/b^2,Q^2)=
-\frac{C_F}{\pi} \int_{b_0^2/b^2}^{Q^2} \frac{dq^2}{q^2} \alpha_s(q^2) \bigg[ \ln\frac{Q^2}{q^2}
-\frac{3}{4} \bigg]\,.
\label{eq::17}
\end{equation}
The first logarithmic term in  eq.~(\ref{eq::17}) is the result of the integration of the soft, singular, part of the $P_{qq}$ splitting function, \textsl{i.e.} the first term on the r.h.s. of eq.~(\ref{eq::9b}).
The term proportional to 3/4 is instead associated to the integration of the regular part
of the splitting function $P_{qq}$, \textsl{i.e.} the second term on the r.h.s. of eq.~(\ref{eq::9b}).
The evaluation of $T_B$, appearing in the third line of eq.~(\ref{eq::14}), is more involved and can be found in the Appendix of Ref.~\cite{KT2}.
We only quote the result
\begin{equation}
T_{B}(Q_0^2,Q^2)=
-\frac{C_F}{\pi} \int_{b_0^2/b^2}^{Q^2} \frac{dq^2}{q^2} (2 \ln \frac{e^{\gamma_E}}{2}) \alpha_s(1/b^2)\,. 
\label{eq::18}
\end{equation}
Adding all terms together the full result reads
\begin{equation}
T_{A+B}(b_0^2/b^2,Q^2)=
-\frac{C_F}{\pi} \int_{b_0^2/b^2}^{Q^2} \frac{dq^2}{q^2}  \bigg[ \alpha_s(q^2) \bigg( \ln\frac{Q^2}{q^2}
-\frac{3}{4} \bigg) +(2 \ln \frac{e^{\gamma_E}}{2}) \alpha_s(1/b^2) \bigg]\,.
\label{eq::19}
\end{equation}
The first term of the integrand produce the leading double logaritmic correction~\cite{PP} and the second and third terms are responsible for part of single logarithmic corrections. Taking into account the one-loop definition of the strong running coupling and the corresponding $\beta$-function, the above integral can be performed analytically.
The results are reported in Refs.~\cite{KT1,KT2}.
As shown in Ref.~\cite{KT2}, the integral of the third term can be conveniently reabsorbed in the integral 
of the two first terms by setting $b_0=2e^{-\gamma_E}$ in the lower integration limit in $T_{A}$. With this replacement 
the final expression of the quark form factor, valid at leading logarithmic accuracy, that we will use for further comparisons reads
\begin{equation}
T_q^{KT}(b_0^2/b^2,Q^2)=
-\frac{C_F}{\pi} \int_{b_0^2/b^2}^{Q^2} \frac{dq^2}{q^2} \alpha_s(q^2) \bigg[ \ln\frac{Q^2}{q^2}
-\frac{3}{4} \bigg]\,.
\label{eq::20}
\end{equation}
As already mentioned, the quark form factor in eq.~(\ref{eq::20}) coincides with the one obtained in Refs.~\cite{KT1,KT2} for TMD fragmentation functions at the same level of accuracy.  
  
\section{AR calculation}
\label{secAR}
\noindent
In this section we consider the results obtained by Aybat and Rogers~\cite{AR}, based on the work 
presented in Ref.~\cite{JC}, and recast their formulas in a form which  should facilitate the comparison with ours. 
In particular we focus on the "B" term of eq.~(26) of that work, which encodes all the perturbative contributions.
The expression for the quark form factor in $b$-space is defined as
\begin{equation}
T^{AR}(\mu_b,\mu,\sqrt{\zeta_F})=
\ln \frac{\sqrt{\zeta_F}}{\mu_b}\tilde{K}(b;\mu_b)+\int_{\mu_b}^{\mu} \frac{d\mu'}{\mu'}
\bigg[ \gamma_F(g(\mu');1)-\ln\frac{\sqrt{\zeta_F}}{\mu'} \gamma_K(g(\mu')) \bigg]\,,
\label{eq::30}
\end{equation}
where $\mu_b=C_1/b$ and the arbitrary constant $C_1$ is set to $C_1=2e^{-\gamma_E} \equiv b_0$. For the present 
purpose, we avoid the introduction of the $b_*$-smoothing prescription. 
We note that TMD distributions do depend on the additional energy parameter 
$\zeta_F$ which results from the TMD PDFs definition elaborated in Ref.~\cite{JC}.
The relevant anomalous dimensions, $\gamma_F$ and $\gamma_K$, and the Collins-Soper kernel can be found in the appendix B of that paper. We report their one-loop expansion here for convenience:
\begin{eqnarray}
\label{eq::31}
\gamma_F(g(\mu);\zeta_F/\mu^2)&=&\alpha_s \frac{C_F}{\pi}\Big(\frac{3}{2}-\ln \frac{\zeta_F}{\mu^2} \Big)+\mathcal{O}(\alpha_s^2)\,,\\
\label{eq::32} 
\gamma_K(g(\mu))&=&2\frac{\alpha_s C_F}{\pi}+\mathcal{O}(\alpha_s^2)\,,\\
\tilde{K}(b,\mu)&=&-\frac{\alpha_s C_F}{\pi} [ \ln(\mu^2 b^2) -\ln 4 +2\gamma_E ]+\mathcal{O}(\alpha_s^2)\,.
\label{eq::33}
\end{eqnarray}
In eq.~(\ref{eq::30}) the factor $\tilde{K}(b,\mu_b)$ evaluated via eq.~(\ref{eq::33}) vanishes. 
This can be seen as the equivalent of setting $Q_0^2=b_0^2/b^2$ in our formalism, see the discussion after eq.~(\ref{eq::15}). Substituting in the expression for the form factor the relevant anoumalous dimensions in eqs.~(\ref{eq::31},\ref{eq::32}) we get
\begin{equation}
T^{AR}(\mu_b,\mu,\sqrt{\zeta_F})=
\int_{\mu_b}^{\mu} \frac{d\mu'}{\mu'}
\bigg[ \alpha_s(\mu') \frac{C_F}{\pi}\frac{3}{2}-\ln\frac{\sqrt{\zeta_F}}{\mu'}  2\frac{\alpha_s(\mu') C_F}{\pi}\bigg]\,.
\label{eq::34}
\end{equation} 
For the purpose of comparison with our formula we change integration variable to 
$q^2=\mu'^2$ with $\mu^2=Q^2$ and $\mu_b=b_0/b$, obtaining
\begin{equation}
T^{AR}(b_0/b,Q^2,\zeta_F)=-\frac{C_F}{\pi}
\int_{b_0^2/b^2}^{Q^2} \frac{dq^2}{q^2} \alpha_s(q^2)
\bigg[ \frac{1}{2} \ln\frac{\zeta_F}{q^2}  -\frac{3}{4}  \bigg]\,.
\label{eq::35}
\end{equation} 
Eq.~(\ref{eq::35}) closely resembles eq.~(\ref{eq::19}). At the level of single logarithms (the 3/4 term) 
the two expressions coincide. However if we set in eq.~(\ref{eq::35}) the natural value  
$\zeta_F=Q^2$ for the energy parameter, we note that the coefficient in front 
of the double logarithmic term is half the one present in the corresponding 
term in  eq.~(\ref{eq::19}).

In order to better understand the origin of this apparent discrepancy 
we may consider the Drell-Yan as a, physical observable, reference process, $q+\bar{q}\rightarrow \gamma^*$.
In the formalism presented in Ref.~\cite{AR}, the soft factor, \textsl{i.e.} the leading double logarithmic
term, can be associated either to the quark or the antiquark depending on the choices
made for the energy parameters $\zeta_q$ and $\zeta_{\bar{q}}$, provided that $\zeta_q \zeta_{\bar{q}}=Q^4$.
This freedom is essentially granted by gauge invariance and the symmetrically sharing 
condition, $\zeta_q=\zeta_{\bar{q}}=Q^2$, is particular appealing since, 
being interested in the evolution of individual TMD PDFs, the quark and antiquark TMD PDFs evolve in the same way.
In our formalism instead the double leading logarithmic term is completely 
associated to the quark line, therefore we would have, in the language of Ref.~\cite{AR}, 
$\zeta_q=Q^4/q^2$ and $\zeta_{\bar{q}}=q^2$. 
Still taking the Drell-Yan process as a reference, this difference essentially arises from aligning the light-like gauge-vector along the antiquark line in the derivation of the quark evolution equations in eq.~(\ref{eq::1}). 
Since with this choice all soft contributions decouple from the antiquark line  (or in AR language $\zeta_{\bar{q}}=q^2)$, the antiquark TMD PDFs 
will evolve with a modified kernel~\cite{KT1,KT2} given by $\widehat{P}_{qq}(u)=C_F[-1-u]_+$ in eq.~(\ref{eq::1}).
The occurance of this kernel may be understood looking at the right hand side 
of eq.~(\ref{eq::9b}). $\widehat{P}_{qq}$ is simply obtained from $P_{qq}$ removing the first, singular, soft term. 
In such an approach~\cite{KT1,KT2}, by inspecting the result in eq.~(\ref{eq::17})
and the following discussion, the  antiquark form factor reads
\begin{equation}
T_{\bar{q}}^{KT}(b_0/b,Q^2)=-\frac{C_F}{\pi}
\int_{b_0^2/b^2}^{Q^2} \frac{dq^2}{q^2} \alpha_s(q^2)
\bigg[-\frac{3}{4}  \bigg]\,.
\label{eq::36}
\end{equation}  
This apparent tension between two formalisms is resolved once gauge independent quantities are evaluated. The form factor needed for the evaluation of the $p_t$-spectrum of the Drell-Yan pair, to leading logarithmic accuracy, reads in fact in both cases  
\begin{equation}
T^{DY,LL}(b_0^2/b^2,Q^2)=T_q(b_0^2/b^2,Q^2)+T_{\bar{q}}(b_0^2/b^2,Q^2)=-\frac{C_F}{\pi}
\int_{b_0^2/b^2}^{Q^2} \frac{dq^2}{q^2} \alpha_s(q^2)
\bigg[ \ln\frac{Q^2}{q^2}  -\frac{3}{2}  \bigg]\,.
\label{eq::37}
\end{equation} 

\section{EIS calculation}
\label{secEIS}
\noindent
Quite recently the issues of factorisation and evolution of TMD distributions 
have been addressed by Echevarria, Idilbi and Scimemi in a number of papers~\cite{EIS1,EIS2,EIS3}. Also in this case it is interesting to compare the results of such a formulation with the ones discussed before. 
The evolution equation for TMD PDFs proposed in Ref.~\cite{EIS2} read
\begin{equation}
F(x,b,Q_f,\mu_f)=F(x,b,Q_i,\mu_i) R(b,Q_i,\mu_i,Q_f,\mu_f)\,,
\end{equation}
where all relevant scales have been explicitely indicated and the evolution 
is again performed in the $b$-space, Fourier-conjugated to the transverse momentum. 
The function $R$ reads
\begin{equation}
\label{Rdef}
R(b,Q_i,\mu_i,Q_f,\mu_f)=\exp\Bigg\{ \int_{\mu_i}^{\mu_f} \frac{d\bar{\mu}}{\bar{\mu}}
\gamma_F \Big(\alpha_s,\ln \frac{Q_f^2}{\bar{\mu}^2} \Big)
\Bigg\} \Bigg( \frac{Q_f^2}{Q_i^2}\Bigg)^{-D(b,\mu_i)}\,.
\end{equation}
We report for convenience the expressions of the various factors appearing in eq.~(\ref{Rdef}):
\begin{eqnarray}
\gamma_F&=&-\frac{1}{2}\bigg[ 2 \Gamma_{\mbox{\tiny{cusp}}} \ln \frac{Q^2}{\bar{\mu}^2}+2\gamma^V \bigg]\,, \\
\Gamma_{\mbox{\tiny{cusp}}} &=& \Gamma_0 \frac{\alpha_s}{4\pi} + \mathcal{O}(\alpha_s^2)\,,\\
\gamma^V&=& \gamma^V_0  \frac{\alpha_s}{4\pi} + \mathcal{O}(\alpha_s^2)\,,\\
D(b,\mu_i)&=&\frac{\Gamma_0}{2} \ln \frac{\mu_i^2 b^2}{4 e^{-2 \gamma_E}}  \frac{\alpha_s(\mu_i)}{\pi}+ \mathcal{O}(\alpha_s^2)\,,
\end{eqnarray}
with $\Gamma_0=4 C_F$ and $\gamma_0^V=-6 C_F$. 
We set for simplicity in eq.~(\ref{Rdef}) $\mu_f=Q_f$ and $\mu_i=Q_i$. Taking logarithms of 
eq.~(\ref{Rdef}) and substituing the above equations we get
\begin{equation}
\label{R2}
\ln R(b,Q_i,Q_f)=-\frac{C_F}{\pi} \int_{Q_i^2}^{Q_f^2} \frac{d\bar{\mu}^2}{\bar{\mu}^2}
\alpha_s(\bar{\mu}^2) \bigg[ \frac{1}{2} \ln \frac{Q_f^2}{\bar{\mu}^2}-\frac{3}{4} \bigg]
-D(b,Q_i) \ln \frac{Q_f^2}{Q_i^2}\,.
\end{equation}
As done in the previous calcualtions, we choose the scale $Q_i^2=b_0^2/b^2$ in order
to remove large logarithms of the type $\ln(Q_i^2 b^2)$. With this setting the $D$ function 
vanishes and we obtain the result
\begin{equation}
\label{R3}
\ln R(b,b_0/b,Q_f)= -\frac{C_F}{\pi} \int_{b_0^2/b^2}^{Q_f^2} \frac{d\bar{\mu}^2}{\bar{\mu}^2}
\alpha_s(\bar{\mu}^2) \bigg[ \frac{1}{2}\ln \frac{Q_f^2}{\bar{\mu}^2}-\frac{3}{4} \bigg]\,,
\end{equation}
which matches the one of AR calculation. It appears therefore that in the EIS formalism 
TMD distributions symmetrically share the soft factor, af fact which in AR language, is translated
to the tacitly assumption $\zeta_q=\zeta_{\bar{q}}=Q_f^2$.
 
\section{Conclusions}
\label{Conc}
\noindent
In this note we have shown that the evolution equation we have proposed some time ago in Ref.~\cite{CT}, 
when a proper limit is taken, can be recast in a form analogous to the results of two indipendent calculations present in the literature. At variance with the latter, the leading logarithmic term contribution to the quark form factor is twice as the ones obtained in the above calculations and reflects the different gauge choices
used in the original derivation of the evolution equations. This way the obtained perturbative form factor
associated to incoming partons is asymmetric. However, as far as physical observables 
are considered, this apparent incosistency is removed and the three formulations give consistent results.

\end{document}